\documentclass[aps,twocolumn,showpacs,amsmath,amssymb,superscriptaddress,prc]{revtex4-2}
\usepackage{graphicx,here}
\usepackage{multirow}
\usepackage{tikz}
\usepackage{epstopdf}
\usepackage[percent]{overpic}
\usepackage{scrextend}
\usepackage{xcolor,xspace}
\usepackage[ulem=normalem]{changes}
\usepackage{hyperref}
\hypersetup{colorlinks=True,urlcolor=blue,linkcolor=blue,citecolor=blue,filecolor=black}
\usepackage{physics}
\usepackage{siunitx}
\usepackage{mathtools}
\usepackage{comment}

\colorlet{Changes@Color}{red}

\usepackage{dcolumn,color,footnote,bm,braket}
\usepackage{url,longtable,tabularx}
\usepackage{threeparttable}

\usepackage{fancybox}
\usetikzlibrary{matrix}

\newcommand\nb{N_\mathrm{B}}
\newcommand\+{\dagger}

\newcommand\hb{\hat{H}_{\mathrm{B}}}
\newcommand\hf{\hat{H}_{\mathrm{F}}}
\newcommand\hbf{\hat{V}_{\mathrm{BF}}}

\newcommand{\be}[4]{B(E2; #1^+_{#2} \!\to\! #3^+_{#4})}
\newcommand{\ben}[4]{B(E2; {#1/2}^-_{#2} \!\to\! {#3/2}^-_{#4})}
\newcommand{\bep}[4]{B(E2; {#1/2}^+_{#2} \!\to\! {#3/2}^+_{#4})}
\newcommand{\bejn}[4]{{#1/2}^-_{#2} \!\to\! {#3/2}^-_{#4}}

\begin{document}

\title{Microscopic formulation
of the interacting boson-fermion model
using the nuclear energy density functional}

\author{M. Homma}
\affiliation{Department of Physics, 
Hokkaido University, Sapporo 060-0810, Japan}

\author{K. Nomura}
\email{nomura@sci.hokudai.ac.jp}
\affiliation{Department of Physics, 
Hokkaido University, Sapporo 060-0810, Japan}
\affiliation{Nuclear Reaction Data Center, 
Hokkaido University, Sapporo 060-0810, Japan}

\date{\today}

\begin{abstract}
Microscopic modeling
of low-energy spectroscopy in medium-heavy
and heavy odd-$A$ nuclei is an outstanding
open problem in nuclear physics.
We propose a novel spectra-generating
collective model for odd-$A$ nuclei
constructed by means of the
nuclear energy density
functional theory and
the interacting
boson-fermion model.
The bosonic Hamiltonian for an
even-even nucleus, which is treated
as a core, and the strength
parameters for the
interactions between the
core and an odd nucleon
are completely determined
by using as microscopic inputs
the potential energy curves
and deformed single-particle spectra
obtained from the self-consistent
mean-field calculations.
In applications to
odd-$A$ Eu, Sm, La, and Ba isotopes,
we demonstrate the validity of the
proposed method in reproducing
reasonably the observed low-energy
spectra and shape phase transitions
in the general cases of the
quadrupole collective
states, that is, nearly spherical,
strongly deformed, and $\gamma$-soft
shapes, in the presence of
an odd nucleon in
a single-$j$ orbit.
\end{abstract}

\maketitle

\section{introduction}\label{sec:introduction}

Microscopic modeling of
nuclei with odd numbers of nucleons
remains a major challenge
for nuclear theory.
In even-even nuclei,
which have even neutron $N$
and proton $Z$ numbers,
to a good approximation
nucleons are coupled pairwise
with spin and parity $J=0^+$,
and this correlation determines,
to a large extent, the nuclear
structure in the vicinity
of the ground state.
As for the odd
(denoted odd-$A$ and odd-odd) nuclei,
one should explicitly take into
account unpaired nucleon degrees
of freedom on the same footing as
the collective degrees of freedom,
and accurately model their couplings
\cite{BM}.

To simplify problems, it is often
reasonable to represent an odd nucleus as a
coupled system of an even-even nucleus
as a core, which is dominated
by collective degrees of freedom,
and unpaired, single nucleons.
Such a framework is provided
by the interacting boson-fermion
model (IBFM) \cite{iachello1979,IBFM},
in which the even-even core
is represented as a many-body
system of bosons in terms of the
interacting boson model (IBM)
\cite{arima1975,IBM},
and certain boson-fermion interactions
are introduced.
The IBFM was applied to
descriptions of
low-energy spectra and electromagnetic
transition properties in
odd-$A$ nuclei in a wide range of the
nuclear chart
(see Refs.~\cite{IBFM-Book,IBFM}
for a review),
to analyze
quantum phase transitions (QPTs)
\cite{jolie2004,cejnar2010,iachello2011,nomura2016qpt,
boyukata2021,fortunato2021}
and shape coexistence
\cite{gavrielov2022,leviatan2025,MAYABARBECHO2025}
in Bose-Fermi systems,
to identify supersymmetric
multiplets
\cite{iachello1980,balantekin1981,jolie2004,frank2009},
and to compute properties of
nuclear decay processes
including single $\beta$
\cite{navratil1988,dellagiacoma1989,yoshida2002,
ferretti2020,nomura2024beta,homma2025beta}
and double $\beta$
\cite{yoshida2013,nomura2022bb,Vsevolodovna2022}
decays.

The basic assumptions of the IBM
are that correlated pairs of
valence nucleons with $J^{\pi}=0^+$
and $2^+$ are associated with
monopole, $s$, and
quadrupole, $d$, bosons,
respectively \cite{OAIT,OAI,IBM}.
While the IBM has been successfully
used for describing the low-energy
collective structure in numerous numbers
of nuclei, strength parameters
of the boson Hamiltonian were
determined to reproduce
the experimental data.
Microscopic formulation of the
IBM, that is, derivation of the
IBM Hamiltonian parameters from
a more fundamental nuclear
structure theory, was also made
by various approaches;
see, e.g.,
Refs.~\cite{OAI,mizusaki1997,nomura2008}.
It was shown, in particular,
that the IBM Hamiltonian for
describing general quadrupole
collective states, i.e.,
those associated with
anharmonic vibrations \cite{nomura2010},
rotations \cite{nomura2011rot},
and $\gamma$-unstable rotations
\cite{nomura2012tri},
was completely determined by using
as microscopic inputs
solutions of the
self-consistent mean-field (SCMF)
calculations that are performed
within the nuclear energy
density functional (EDF) framework
\cite{RS,bender2003,vretenar2005,robledo2019}.
This method was extensively
used in nuclear structure studies
describing nuclear spectroscopy
related to, e.g., shape coexistence,
octupole and hexadecapole shapes,
in even-even systems
(see the review \cite{nomura2025rev} and
references therein).

The microscopic foundation of the IBFM
has also been investigated,
e.g., in terms of the generalized
seniority scheme of the nuclear shell model
\cite{scholten1985,scholten1981boson,
otsuka1987,yoshinaga2000}.
The EDF-based IBM, mentioned above,
was extended to compute the spectroscopy
of odd-$A$ nuclei
in Ref.~\cite{nomura2016odd}.
In this method \cite{nomura2016odd},
the EDF-SCMF calculations
yield the single-particle
energies and occupation probabilities
for an odd particle,
which are used to construct
the IBFM Hamiltonian.
However,
coupling constants for the
boson-fermion interaction terms
were left as free parameters,
and were determined to reproduce
to a certain accuracy
the observed low-energy spectrum
in each odd-$A$ nucleus.

In the preceding work
\cite{homma2025plb},
we proposed a method to determine
completely the IBFM Hamiltonian
parameters by the EDF calculations.
In the first step,
the IBM Hamiltonian describing the
even-even core nucleus
was determined by employing the
method of Refs.~\cite{nomura2008,nomura2010},
that is, the EDF-SCMF potential energy
curve (PEC) along the quadrupole
deformation $\beta$ is mapped onto
the expectation value of the
IBM Hamiltonian in the condensate
state of $s$ and $d$ bosons.
In the second step,
the boson-fermion strength
parameters were derived so that
the deformed single-particle
spectra as functions of the
axial deformation $\beta$
in the intrinsic frame
of the IBFM
\cite{IBFM,leviatan1988,leviatan1989,
iachello2011,petrellis2011}
were made to match those
of the EDF-SCMF
calculations for odd-$A$ nuclei.
In an illustrative application
to axially deformed $^{149-159}$Eu
isotopes, using the single-$j$
(i.e., proton $1h_{11/2}$) orbit,
this method was shown to be valid
in reproducing excitation energies of
low-energy negative-parity states
without any adjustment of parameters
to the experimental data.
The calculation, in particular,
correctly reproduced the change of the
angular momentum $I$ of the lowest-energy
negative-parity states from $I={11/2}^-$
to ${5/2}^-$ near the neutron
number $N=90$,
which can be considered to be
an empirical signature
of the shape QPTs
in the odd-$A$ systems
\cite{iachello2011,petrellis2011}.

The results reported in
Ref.~\cite{homma2025plb} demonstrated
the first successful case of the
IBFM description with all the
model parameters derived from the
microscopic EDF inputs.
From a practical point of view,
the method developed in
\cite{homma2025plb}
is also considered to be an
alternative EDF-based collective model
able to provide predictions
for low-energy spectroscopy
in heavy odd-$A$ nuclei.
One might rather consider that
a straightforward way to compute
spectroscopic properties in
the EDF model would be
to project the mean-field
solutions onto states
with good symmetry quantum numbers,
and to incorporate quantum fluctuation
effects via configuration mixing
by means of the
generator coordinate method (GCM).
The GCM was actually implemented
in the EDF to study spectroscopy in
odd-$A$ nuclei, by fully taking
into account
the breaking of the time-reversal
invariance and blocking effects
\cite{bally2014,borrajo2016,zhou2024}.
However, practical applications
of the projected EDF-GCM approach
to a large number of odd-$A$ and odd-odd nuclei
are prohibitively demanding,
and have been limited to
light odd-$A$ nuclei,
e.g., Mg isotopes.
In recent years advances have also
been made in the
{\it ab-initio} GCM descriptions
of odd-$A$ nuclei that are
heavier than Mg
(e.g., Ref.~\cite{li2025}).
These present a promising microscopic
approach to odd-$A$ and odd-odd
systems in medium to heavy
mass regions.

Following up the initial
study of Ref.~\cite{homma2025plb},
this article reports more systematic
spectroscopic calculations for odd-$A$ nuclei
within the EDF-IBFM approach.
The scope of the present paper is
(i) to address the validity of the
proposed method to describe,
in addition to axially deformed systems,
axially asymmetric nuclei
that are expected to show
pronounced $\gamma$ softness;
(ii) to show the robustness of
the method in describing reasonably
the odd-$N$ and odd-$Z$ systems,
in which the odd nucleon
i.e., neutron and proton,
respectively,
is the same as and different
from the control parameter of the
QPT (nucleon number)
along the isotopic chain;
and (iii) to develop a way of
calculating the electromagnetic
transition properties without
any adjustment of the
effective boson charges.

Nuclei we specifically consider
in this study are
axially deformed even-even $^{148-158}$Sm and
odd-$A$ $^{149-159}$Sm and $^{149-159}$Eu
isotopes near the
transitional region $N\approx90$,
and $\gamma$-soft
even-even $^{128-134}$Ba and
odd-$A$ $^{129-135}$La
and $^{127-133}$Ba isotopes
in the mass $A\approx130$ region.
In Ref.~\cite{homma2025plb},
we discussed the shape
transitions in odd-$A$ Eu,
which correspond to the manifest
QPTs that are suggested to
occur in the neighboring even-even Sm
isotopes from the U(5)
(associated with spherical vibrational shapes)
to the SU(3)
(strongly prolate deformed shapes)
dynamical symmetries.
The additions of the results
for the Ba and La nuclei
are used to investigate whether the
proposed method works reasonably
well in $\gamma$-soft systems,
as in the case of the axially-deformed
Eu and Sm.
In the considered La and Ba region,
in particular, another type of
shape change between
the U(5) and O(6) (representing
$\gamma$-unstable shapes)
dynamical symmetries is
expected to occur
\cite{IBM,iachello2000-E5,casten2000-E5}.

The paper is organized as follows.
In Sec.~\ref{sec:model},
we introduce the procedure 
to derive the IBFM Hamiltonian
parameters from the EDF-SCMF
calculations.
Section~\ref{sec:results}
provides calculated results,
including the IBFM and HFB
deformation energy curves,
deformed single-particle
energies, excitation energies
for low-lying states,
and electric quadrupole ($E2$)
transitions.
A summary is given and
insights into possible future
studies are provided in
Sec.~\ref{sec:summary}.

\section{theoretical framework}\label{sec:model}

In the following,
we use the simplest versions
of the IBM and IBFM, in which
like-neutron and like-proton
bosons are not distinguished.
The total number of bosons,
denoted $\nb$, equals the number
of pairs of valence nucleons
with respect to the nearest
doubly magic nucleus.
For the even-even
$^{148-158}$Sm, $\nb=8,9,\ldots,13$,
respectively, while for
the even-even $^{128-134}$Ba,
$\nb=8$, 7, 6 and 5, respectively,
as the neutrons are hole-like.
The IBFM Hamiltonian consists
of the bosonic (IBM) Hamiltonian,
$\hb$, single-nucleon Hamiltonian
for an odd particle,
$\hf$, and the particle-core
interaction, $\hbf$ \cite{IBFM}:
\begin{equation}
\hat{H} = \hb + \hf + \hbf .
\label{eq:ham}
\end{equation}
The IBM Hamiltonian
of the form
\begin{equation}
\hb = \epsilon_d\hat{n}_d + \kappa\hat{Q}\cdot\hat{Q}
\label{eq:hb}
\end{equation}
is employed,
where the first
term stands for the
$d$-boson number operator,
$\hat{n}_d = d^\+\cdot\tilde{d}$,
with $\epsilon_d$ being
the single-$d$-boson energy,
and the second term
the quadrupole-quadrupole
boson interaction with
the strength $\kappa$.
The quadrupole operator
is given as
$\hat{Q} = s^\+\tilde{d}+d^\+\tilde{s}
+\chi[d^\+\times\tilde{d}]^{(2)}$,
with $\chi$ being another parameter.
Note the notations $\tilde{s} = s$ and 
$\tilde{d}_\mu = (-1)^\mu d_{-\mu}$.

As in the previous work
\cite{homma2025plb},
we shall consider single-$j$ problems,
that is, for the single-particle
space we take the proton $1h_{11/2}$
and neutron $1i_{13/2}$ orbits
for the odd-$A$ Eu and Sm, respectively,
and the neutron and proton $1h_{11/2}$
orbits for the odd-$A$ Ba and La,
respectively.
The spectroscopic properties
in the odd-$A$ nuclei
discussed in the following, therefore,
all arise from
the configurations of
the unique-parity orbit
coupled to the even-even boson core.
Since the scope of this study
includes addressing
the validity of the procedure
in a systematic way
in different mass regions,
the assumption of taking into account
only a single-$j$ orbit for
the odd particle would not lose
much generality.
The extension to the multiple-$j$
cases can be straightforward,
but is also beyond the scope of the
present study.

In the single-$j$ case,
the Hamiltonian $\hf$ in
\eqref{eq:ham} is given simply as 
\begin{equation}
\hf = -\epsilon_j\sqrt{2j + 1}
[a_j^\+ \cross \tilde{a}_j]^{(0)}
\equiv \epsilon_j\hat{n}_j
\; ,
\label{eq:hf}
\end{equation}
where $a_j^{\+}$ and
$\tilde{a}_{j,m} \equiv (-1)^{j-m}a_{j,-m}$
stand for the particle creation and
annihilation operators, respectively,
and $\epsilon_j$ is the
single-particle energy.

The boson-fermion interaction,
$\hbf$, can be expressed in the
following compact form, consisting
only of three essential interactions
that are sufficient for numerical studies
\cite{IBFM}:
\begin{equation}
\hbf = \Gamma\hat{V}_\mathrm{dyn}
+ \Lambda\hat{V}_\mathrm{exc}
+ A\hat{V}_\mathrm{mon} \; ,
\label{eq:hbf}
\end{equation}
where the first, second, and
third terms are, respectively,
so-called dynamical
(quadrupole) term, representing
direct boson-fermion interactions,
the exchange term, which reflects that
bosons are made of nucleon pairs,
and the monopole interaction.
They are expressed as
\begin{align}
&\hat{V}_\mathrm{dyn}
= \hat{Q}\cdot[a_j^\+\cross\tilde{a}_j]^{(2)}
\label{eq:dyn} \\
&\hat{V}_\mathrm{exc}
= :[[d^\+\cross\tilde{a}_j]^{(j)}
\times[\tilde{d}\cross a_j^\+]^{(j)}]^{(0)}:
\label{eq:exc} \\
&\hat{V}_\mathrm{mon}
= \hat{n}_d\hat{n}_j
\label{eq:mon} \; .
\end{align}
The notation $:[\cdots]:$
in Eq.~\eqref{eq:exc} denotes
normal ordering.
In the generalized seniority scheme,
the strength parameters 
$\Gamma$, $\Lambda$, and $A$
in Eq.~\eqref{eq:hbf} are further
simplified to be of the forms
\cite{scholten1985}
\begin{align}
&\Gamma = \Gamma_0\gamma_{jj}
\label{eq:coef_dyn}\\
&\Lambda = -2\Lambda_0\sqrt{\frac{5}{2j+1}}\beta_{jj}^2
\label{eq:coef_exc} \\
&A = A_0
\label{eq:coef_mon}
\; ,
\end{align}
where $\gamma_{jj} = (u_j^2-v_j^2)Q_{jj}$
and $\beta_{jj} = 2u_jv_jQ_{jj}$, 
with $Q_{jj} = \mel{j}{|Y^{(2)}|}{j}$
being the matrix element of the
quadrupole operator 
in the single-particle basis.
$v_j$ and $u_j$ are the occupation
and unoccupation amplitudes
for the orbital $j$,
satisfying the relation
$u^2_j+v^2_j=1$.
$\Gamma_0$, $\Lambda_0$, and $A_0$
are strength parameters.

To associate with the intrinsic
properties generated by the SCMF method,
the geometry of the IBFM Hamiltonian
\eqref{eq:ham} is formulated
using the following basis:
\begin{equation}
\ket{\nb;\bar{\beta},\bar{\gamma};j,m}
= \ket{\nb;\bar{\beta},\bar{\gamma}}\otimes\ket{j,m}\; .
\label{eq:coherent}
\end{equation}
$\ket{j,m}$ stands for the
single-particle basis for an odd nucleon,
$\ket{j,m} = a^\+_{j,m}\ket{0}$.
$\ket{\nb;\bar{\beta},\bar{\gamma}}$
denotes the coherent state \cite{ginocchio1980}
of $s$ and $d$ bosons, given as
\begin{equation}
\ket{\nb;\bar{\beta},\bar{\gamma}} = (\nb!)^{-1/2}(b_c^\+)^{\nb}\ket{0}
\; ,
\label{eq:coherent_b}
\end{equation}
with
\begin{equation}
b_c^\+ =
(1+\bar{\beta}^2)^{-1/2}
\qty[s^\+ + d_0^\+\bar{\beta}\cos{\bar{\gamma}}
+\frac{1}{\sqrt{2}}(d_{+2}^\+
+d_{-2}^\+)\bar{\beta}\sin{\bar{\gamma}}] \; .
\label{eq:coherent_op}
\end{equation}
$\bar{\beta}$ and
$\bar{\gamma}$ are boson analogs
of the axial quadrupole
deformation $\beta$ 
and triaxiality $\gamma$ \cite{BM},
respectively. 
$\ket{0}$ represents the inert core
or a doubly magic nucleus, i.e.,
$^{132}$Sn.
For simplicity,
we assume that the bosonic
deformation $\bar{\beta}$ is
proportional to the fermionic
deformation $\beta$
\cite{ginocchio1980,nomura2008}, 
and that the bosonic triaxiality
$\bar{\gamma}$ is the same
angle variable as the fermionic
counterpart $\gamma$:
\begin{align}
&\bar{\beta} = C_\mathrm{B}\beta
\label{eq:beta}
\\
&\bar{\gamma} = \gamma
\label{eq:gamma}
\end{align}
with $C_\mathrm{B}$
being a constant of proportionality.
The factor $C_{\rm B}$ in \eqref{eq:beta}
is introduced to take into
account the difference in
model space between the IBM
and SCMF model:
the former comprises valence
nucleons in a given model space,
while in the latter
all constituent nucleons
are considered in a much
larger fermionic configuration space.
By taking the expectation value of the
IBFM Hamiltonian \eqref{eq:ham}
in the basis \eqref{eq:coherent},
the energy surface for an odd-$A$
nucleus is obtained in a
matrix form \cite{leviatan1988}.
In this study, we assume axial
symmetry ($\gamma = \ang{0}$),
and hereafter omit the
argument $\gamma$.
This assumption makes
the energy surface matrix
diagonal, and each diagonal element,
given by
\begin{equation}
E_K = E_\mathrm{B}(\beta;\xi) + \lambda_K(\beta;\eta)
\; ,
\label{eq:IBFM_PES}
\end{equation}
corresponds
to the energy of the state with
the projection $m = K$.
The entire energy surface for
the IBFM is calculated as the sum of
the bosonic energy surface,
$E_\mathrm{B}(\beta;\xi)$,
and the single-particle energy,
$\lambda_K(\beta;\eta)$.
Here $\xi$ and $\eta$ denote
sets of the parameters involved in
$E_{\rm B}$ and $\lambda_{K}$,
respectively:
\begin{align}
\xi = \{\epsilon_d,\kappa,\chi,C_\mathrm{B}\}
\, ,
\quad
\eta = \{\Gamma_0,\Lambda_0,A_0\}
\; .
\end{align}
Each term in \eqref{eq:IBFM_PES}
is calculated as \cite{leviatan1988,IBFM}
\begin{align}
E_\mathrm{B}(\beta;\xi)
=&
\ev*{\hb}{\nb;\bar{\beta}}
\nonumber\\
=&
\frac{\nb}{1+\bar{\beta}^2}
\qty{5\kappa+\qty[\epsilon_d+\kappa(1+\chi^2)]\bar{\beta}^2}
\nonumber \\
&+ \kappa \frac{\nb(\nb-1)\bar{\beta}^2}{(1+\bar{\beta}^2)^2}
\qty(2 - \chi\sqrt{\frac{2}{7}}\bar{\beta})^2
\label{eq:IBM_PES} 
\\
\lambda_K(\beta;\eta)
=&
\ev*{(\hf+\hbf)}{\nb;\bar{\beta};j,m}
\nonumber\\
=& \epsilon_j + A\frac{\nb\bar{\beta}^2}{1+\bar{\beta}^2} + 
\frac{\nb\bar{\beta}}{1+\bar{\beta}^2}\qty[3K^2-j(j-1)]P_j
\nonumber \\
\times\Bigg\{\Gamma\Bigg(\chi
&\sqrt{\frac{2}{7}}\bar{\beta}-2\Bigg)-\Lambda P_j\sqrt{2j+1}
[3K^2-j(j+1)]\bar{\beta}\Bigg\} , 
\label{eq:spe}
\end{align}
where $P_j = [(2j-1)j(2j+1)(j+1)(2j+3)]^{-1/2}$.
We assume that $\chi$ in
Eq.~\eqref{eq:spe} is 
the same as that in
Eq.~\eqref{eq:IBM_PES}.

The model parameters,
$\xi$ for the IBM cores
and $\eta$
for the boson-fermion interaction,
are determined in the following way
\cite{homma2025plb}.
\begin{enumerate}
\item
In the first step,
we perform the constrained SCMF
calculations to obtain
the one-dimensional PEC
for even-even core nuclei
and deformed single-particle energies
for the neighboring odd-$A$ nuclei.
The constraints to the SCMF
calculations are those on the mass
quadrupole moment $Q_{20}$,
which is related to the
geometrical deformation $\beta$.
In this study, we adopt the
Hartree-Fock-Bogoliubov (HFB)
method using the SkM* interaction
\cite{skms} of the Skyrme EDF
\cite{Skyrme}.
The SCMF calculations
are carried out by using
the computer program HFBTHO
(v4.0) \cite{hfbtho400}.

\item
Next, we determine the parameters
for the boson-core Hamiltonian
$\hb$ by mapping 
the HFB PEC obtained
in the previous step
onto the IBM energy
curve \eqref{eq:IBM_PES}:
\begin{align}
 E_{\rm B}(\beta;\xi) \approx
E_{\rm HFB}(\beta)
\; .
\end{align}
In this procedure,
the set of the IBM parameters,
$\xi$, are calibrated so that basic
characteristics of the bosonic
energy curves in the vicinity of
the equilibrium minimum,
such as the depth of the potential,
and curvature up to a few MeVs
from the minimum,
should be made similar to
those of the HFB PEC.
The scale factor $C_\mathrm{B}$
\eqref{eq:beta} is determined
so that IBM PEC reproduces the
location of the energy minimum
in the HFB PEC.

\item
In the final step,
the strength parameters, $\eta$,
for the boson-fermion interaction,
$\hbf$ \eqref{eq:hbf},
are fixed so that the single-particle
energies of the IBFM, $\lambda_K(\beta;\eta)$
\eqref{eq:spe}, should reproduce
for each $K$ value
the HFB counterparts,
$\epsilon_K(\beta)$, near the
equilibrium minimum:
\begin{align}
\label{eq:lambda}
 \lambda_K(\beta_e;\eta)\approx\epsilon_K(\beta_e)
\; .
\end{align}
By this procedure, the behaviors of
the IBFM deformed single-particle
spectra are made to resemble those
of the HFB ones within the
range of the deformation
$0\leqslant\beta\leqslant\beta_e$,
where $\beta_e$ is the deformation
corresponding to the minimum 
in the HFB PEC for an odd-$A$ nucleus.
In the HFB calculations
the blocking effects are taken into
account at each deformation
$\beta$ and for all possible
single-particle orbits \cite{hfbtho400}.
The $v_j$ and $u_j$ amplitudes,
which enter the formulas
\eqref{eq:coef_dyn} and
\eqref{eq:coef_exc},
are obtained from the
Skyrme-HFB calculations performed
at the spherical configuration,
i.e., those that are
constrained to zero
quadrupole deformation,
as was done in Ref.~\cite{nomura2016odd}.
\end{enumerate}
The IBFM Hamiltonian
\eqref{eq:ham}
with all the parameters fixed
by the aforementioned procedure
is numerically diagonalized
in the basis $\ket{L\otimes j;I}$
\cite{NPBOS},
where $L$ denotes the angular
momentum of the boson system
and $I$ the total angular momentum
of the odd-$A$ system.
This gives rise to energy levels,
and wave functions to compute
electromagnetic transition
properties.

\section{Results}\label{sec:results}

\subsection{Potential energy curves
of even-even nuclei}\label{subsec:pes-e}

\begin{figure}[htbp]
\centering
\includegraphics[width=\linewidth]{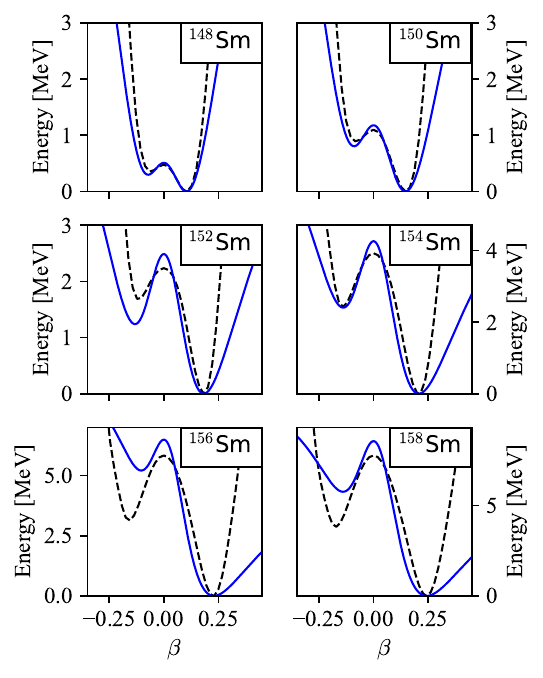}
\caption{Potential energy curves along
the axial quadrupole deformation $\beta$
for the even-even nuclei $^{148-158}$Sm
calculated within the Skyrme-HFB (dashed curves)
and IBM (solid curves).}
\label{fig:pec-evensm}
\end{figure}

\begin{figure}[htbp]
\centering
\includegraphics[width=\linewidth]{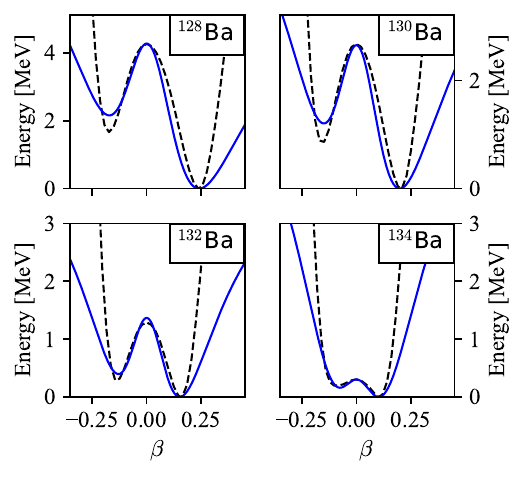}
\caption{Same as the caption to
Fig.~\ref{fig:pec-evensm}, but for
$^{128-134}$Ba.}
\label{fig:pec-evenba}
\end{figure}

In Figs.~\ref{fig:pec-evensm} and
\ref{fig:pec-evenba},
the IBM PECs for the
even-even-core nuclei are
shown, together with
the original Skyrme-HFB PECs.
The HFB PECs for the Sm
isotopes show patterns of
an empirically
suggested shape phase transition,
that is,
a transition from a weakly
deformed prolate shape
with an equilibrium minimum
$\beta_e = 0.10$
in $^{148}$Sm to a strongly deformed
prolate shape near $^{154}$Sm.
The IBM PECs appear to
reproduce overall behaviors of
the Skyrme-HFB PECs.
However, as the deformation becomes
larger, $\beta>0.3$,
the IBM PEC becomes flat
compared with the HFB one.
This deviation reflects
the difference in the
size of the model space and
degrees of freedom
between the IBM and HFB,
as pointed out in Sec.~\ref{sec:model}.
In the HFB configurations
of the large deformations,
$\beta\gg\beta_{e}$,
noncollective, quasiparticle excitations
begin to play a part, but these
modes of excitations are outside of
the standard IBM model space.
This is the reason why the
mapping is made to reproduce
features of the HFB PEC
in the vicinity of the equilibrium
minimum.
One also notices a discrepancy
between the HFB and IBM curves
near the oblate saddle points
in heavier Sm, most notably,
in $^{156}$Sm and $^{158}$Sm.
This arises because the PEC
mapping is made for the
energy range of up to 2-3 MeV from
the equilibrium minimum.
In the above two nuclei,
in which the potential is 
particularly deep,
the IBM reasonably
reproduces the behavior of the HFB PEC
on the prolate side,
but this is not the case
with the oblate side.

The HFB PECs shown in Fig.~\ref{fig:pec-evenba}
suggest a prolate
shape in $^{128}$Ba and $^{130}$Ba.
The energy difference
between the prolate equilibrium
minimum at $\beta_e\approx0.2$
and the oblate saddle point
becomes smaller as one approaches
$^{134}$Ba.
This implies that the potential
becomes soft in $\gamma$ deformation,
and thus indicates a
prolate-to-$\gamma$-soft shape transition.
It should be noted that,
while the present study
is restricted to axially symmetric
deformation, the triaxiality
would have potential impacts
on the IBM parameters and
resulting low-lying states.
In fact, earlier
SCMF calculations of the
triaxial quadrupole
deformation energy surfaces that are
based on the nonrelativistic
Skyrme (e.g., \cite{nomura2010}), Gogny
(e.g., \cite{delaroche2010,nomura2017odd-3}),
and relativistic
(e.g., \cite{li2010,nomura2017odd-1})
EDFs suggested
a shallow triaxial minimum
with a nonzero $\gamma$
deformation in the nuclei
in the $A\approx130$ Ba region.
The $\gamma$ softness is
here accounted for in the IBM
by the value of the parameter
$\chi$ that is small in magnitude,
since in the $\gamma$-unstable
O(6) limit of the IBM this parameter
is supposed to be zero.
To describe more quantitatively
possible effects of the triaxiality
on low-lying states,
cubic (three-body) boson terms
would need to be included in the IBM
Hamiltonian, with their strength
parameters being obtained from
mapping of the triaxial HFB energy
surfaces \cite{nomura2012tri}.
This requires a major extension
of the method, but would be an
interesting future work.

\subsection{Deformed single-particle spectra}\label{subsec:spe}

\begin{figure*}[htbp]
\centering
\includegraphics[width=\linewidth]{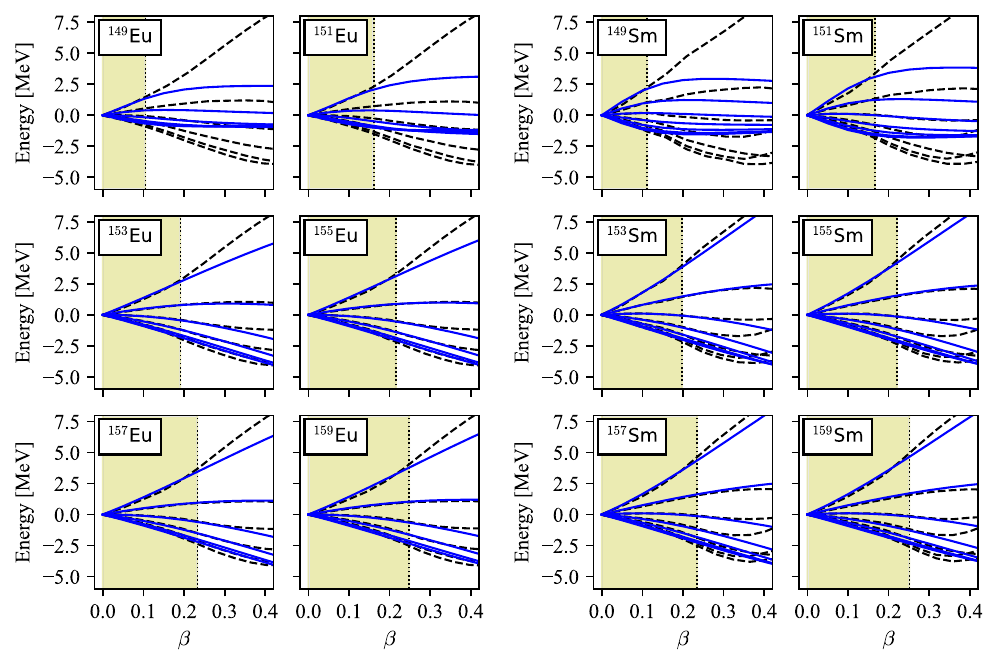}
\caption{Deformed single-particle
spectra calculated for the odd-$Z$ nuclei
$^{149-159}$Eu and odd-$N$ nuclei
$^{149-159}$Sm within the IBFM
(solid curves), Eq.~\eqref{eq:spe},
and the HFB-SCMF method
(dashed curves).
Yellow-shaded areas indicate
the range of the deformation
$\beta$, within which the IBFM
single-particle spectra are
matched to those of the HFB.
The IBM single-particle spectra
corresponding to different values
of the $K$ quantum number,
$K=1/2$, 3/2, 5/2, 7/2, 9/2, and 11/2
($K=1/2$, 3/2, 5/2, 7/2, 9/2, 11/2, and 13/2),
in increasing order in energy,
are compared with the HFB ones,
which are labeled by the asymptotic
quantum numbers $K[Nn_zm_l]$
that are equal to $1/2[550]$,
$3/2[541]$, $5/2[532]$, $7/2[523]$,
$9/2[514]$, $11/2[505]$
($1/2[660]$,
$3/2[651]$, $5/2[642]$, $7/2[633]$,
$9/2[624]$, $11/2[615]$, $13/2[606]$),
respectively, for the odd-$A$ Eu (Sm)
isotopes}
\label{fig:spe-eusm}
\end{figure*}

\begin{figure*}[htbp]
\centering
\includegraphics[width=\linewidth]{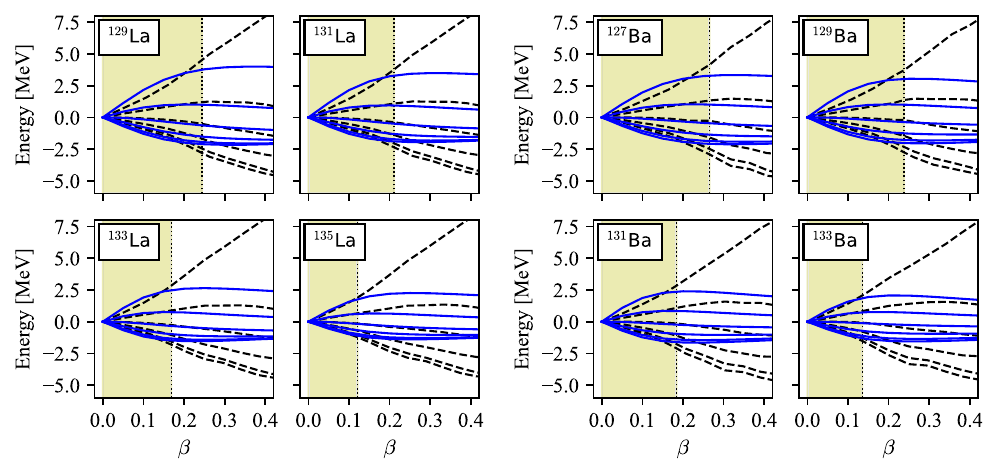}
\caption{Same as the caption to
Fig.~\ref{fig:spe-eusm},
but for the odd-$Z$
$^{129-135}$La and odd-$N$ $^{127-133}$Ba.}
\label{fig:spe-laba}
\end{figure*}

Figure~\ref{fig:spe-eusm}
depicts the IBFM and Skyrme-HFB
deformed single-particle energies
for the odd-$Z$ nuclei, $^{149-159}$Eu,
originating from the proton $1h_{11/2}$
spherical orbit,
and those for the odd-$N$ nuclei,
$^{149-159}$Sm, originating
from the neutron $1i_{13/2}$ orbit.
It is seen that the IBFM single-particle
orbits $\lambda_K(\beta;\eta)$
reproduce reasonably well the
HFB counterparts within the ranges
$0\leqslant\beta\leqslant\beta_e$.

One may also notice that,
for very large quadrupole deformations
($\beta\gg\beta_e$),
the IBFM single-particle energies
become flat, whereas those from the
HFB calculations
in general exhibit much more significant
changes with $\beta$.
This difference arises
from the limited IBFM model space
as compared with that of the HFB,
and we have seen in Fig.~\ref{fig:pec-evensm}
a similar discrepancy
when comparing the HFB and IBM PECs.
As pointed out earlier,
the SCMF configurations at the
very large deformations are
beyond the model spaces of the
standard IBM and IBFM, hence
the comparisons of the IBFM with the
HFB single-particle orbits
in the region $\beta\gg\beta_e$
would not make much sense,
but should be made within the
range $0\leqslant\beta\leqslant\beta_e$.

The relation \eqref{eq:lambda}
used to obtain the set of the IBFM parameters,
$\eta$, turns out to be a good
approximation for those odd-$A$
nuclei with $N\leqslant89$,
which are moderately deformed,
$\beta_e\approx0.15$.
For these nuclei,
the IBFM single-particle energies
reasonably reproduce the HFB ones
in the range
$0\leqslant\beta\leqslant\beta_e$.
However, for more deformed odd-$A$ nuclei
with $N\geqslant90$,
which typically exhibit a larger
$\beta_e$ value, $\beta_e\geqslant0.2$,
the relation of \eqref{eq:lambda}
does not appear to hold,
because the behaviors of
the IBFM energies,
$\lambda_{K}(\beta;\eta)$,
differ significantly from those
of the HFB ones near
the configurations
$\beta\approx\beta_e$.
The reasons for this to occur are,
as mentioned, the limited size of and degrees
of freedom in the IBFM configuration
space, and also the fact that
the analytical form of the formula
\eqref{eq:spe} may be
of too simplified a form.
To take these into account,
in Ref.~\cite{homma2025plb}
we assumed that, specifically
for well deformed odd-$A$ Eu
nuclei typically exhibiting
$\beta_e\geqslant0.2$,
the scale factor $C_{\rm B}$
for the $\beta$ deformation for the
bosonic part, $E_{\rm B}(\beta;\xi)$
\eqref{eq:IBM_PES}, could be
different from that in the
boson-fermion coupling part,
$\lambda_{K}(\beta;\eta)$
\eqref{eq:spe},
and we replaced $C_{\rm B}$
in the formula
\eqref{eq:spe} with a
new parameter $C_\mathrm{BF}$.
We here apply this procedure to the
odd-$A$ Sm nuclei as well,
and consider the following formula
for the deformed odd-$A$ Eu nuclei
with $N\geqslant90$
and Sm with $N\geqslant91$:
\begin{equation}
\label{eq:gen_pec}
E_K(\beta) = E_\mathrm{B}(\beta;\xi) + \lambda_K(\beta;\eta')
\; ,
\end{equation}
where $\eta'$ denotes
the set of parameters,
$\eta' = \{A_0,\Gamma_0,\Lambda_0,C_\mathrm{BF}\}$.
For those Eu and Sm isotopes
with $N\leqslant88$ and $N\leqslant90$,
respectively, common $C_{\rm B}$
values are considered for both the
boson, $E_{\rm B}(\beta;\xi)$,
and boson-fermion, $\lambda_{K}(\beta;\eta)$,
parts, i.e., $C_{\rm B}=C_{\rm BF}$.
Using different scale factors,
i.e., $C_{\rm B}$ for boson part
and $C_{\rm BF}$ for boson-fermion
interaction part,
for those nuclei with $N\geqslant 90$,
effectively takes into account
the fact that,
while the single-particle orbit
couples weakly to the even-even
core that is close to nearly spherical
vibrational limit,
in strongly deformed systems
the particle-core coupling
is expected to be stronger.
Specifically, the formula
\eqref{eq:gen_pec} is used in those
cases in which the HFB energy curves
exhibit a substantially steep ($>2$ MeV)
potential with a distinct axial prolate
minimum typically found
at $\beta_e\geqslant 0.2$,
such as the even-even
Sm nuclei with $A\geqslant 152$
(see Fig.~\ref{fig:pec-evensm}).

Figure~\ref{fig:spe-laba}
shows the HFB and IBFM deformed
single-particle spectra for $^{129-135}$La
and $^{127-133}$Ba, which originate
from the proton and neutron $1h_{11/2}$
spherical orbits, respectively.
One can observe that
the IBFM reasonably
reproduces the HFB
single-particle energies within
the ranges
$0\leqslant\beta\leqslant\beta_e$.
For these odd-$A$ nuclei,
the formula \eqref{eq:IBFM_PES}
is consistently used
for all the La and Ba
isotopes considered.
This is mainly because
the quadrupole correlations
in these systems are, in general,
not as pronounced
as in the case of the Eu and Sm nuclei.
That is, even though a large quadrupole
deformation, $\beta_e\approx0.2$,
is suggested for some of these odd-$A$
La and Ba nuclei,
as compared with those
for the Sm and Eu nuclei,
the HFB PECs for the corresponding
even-even Ba nuclei exhibit
potentials that are rather
shallow and are expected to
be soft in $\gamma$ deformation,
as the energy difference
between the prolate equilibrium
minimum and the oblate saddle point
is more or less small.

\begin{figure*}[htbp]
\centering
\includegraphics[width=\linewidth]{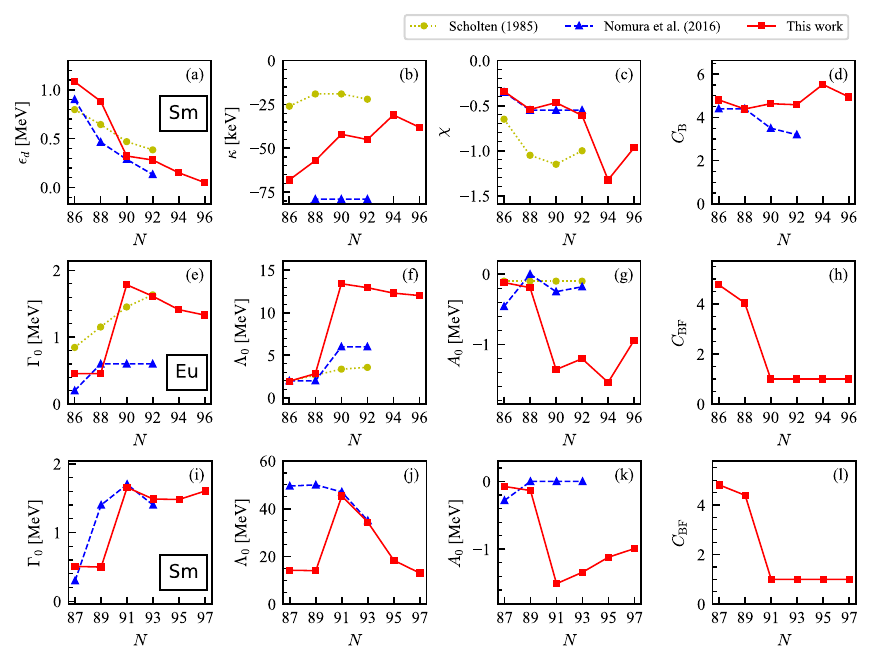}
\caption{
Derived IBM parameters
for the even-even $^{148-158}$Sm
[(a)--(d)], boson-fermion interaction
strengths and $C_{\rm BF}$
values for the
odd-$A$ $^{149-159}$Eu [(e)--(h)]
and $^{149-159}$Sm [(i)--(l)]
(represented by the squares
connected by the solid curves).
The values of the parameters
denoted ``This work'' in (a)--(h) are
adopted from Ref.~\cite{homma2025plb}.
Also in (a)--(c),
in (e)--(g), and in (i)--(k),
the IBM and IBFM parameters used in
Ref.~\cite{nomura2016qpt}
[``Nomura {\it et al.} (2016),''
represented by the solid triangles] and
Ref.~\cite{scholten1985}
[``Scholten (1985),'' solid circles]
are shown.}
\label{fig:para-eusm}
\end{figure*}

\begin{figure*}[htbp]
\centering
\includegraphics[width=\linewidth]{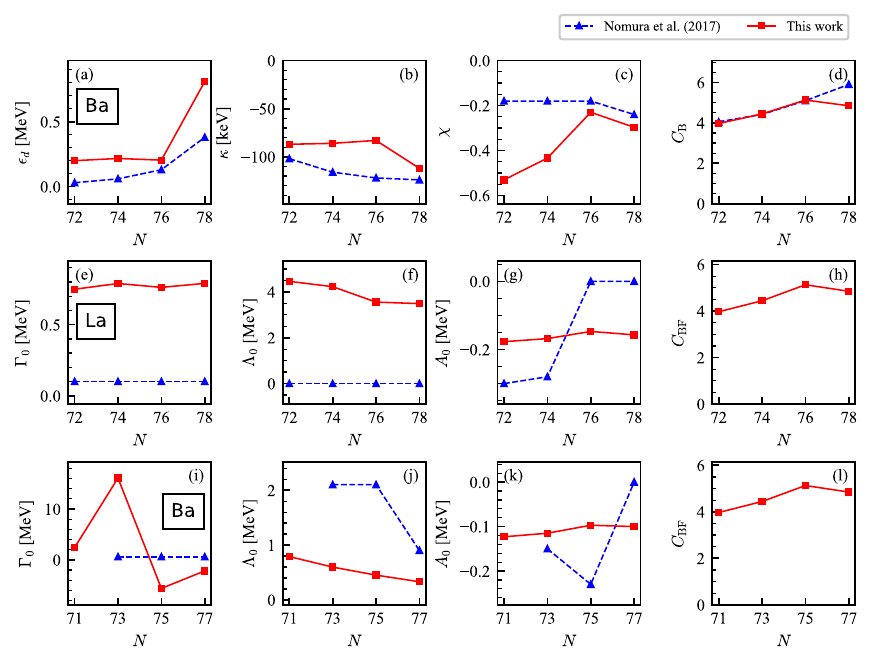}
\caption{
Similar to the caption to
Fig.~\ref{fig:para-eusm},
but for those parameters obtained
in the present work for even-even $^{128-134}$Ba
[(a)--(d)], odd-$Z$ $^{129-135}$La [(e)--(h)],
and odd-$N$ $^{127-133}$Ba
[(i)--(l)].
The parameters used in the
previous IBFM study of
Ref.~\cite{nomura2017odd-1}
are also shown [``Nomura {\it et al.} (2017),'' triangles]}
\label{fig:para-laba}
\end{figure*}

\subsection{Derived parameters}\label{subsec:para}

The derived IBFM parameters
for all the considered odd-$A$ isotopes
are presented in Figs.~\ref{fig:para-eusm}
and \ref{fig:para-laba}.
Table~\ref{tab:vv} summarizes
the occupation probabilities for the
spherical single-particle orbits.
The parameters
for the even-even Sm and odd-$A$ Eu,
plotted in Figs.~\ref{fig:para-eusm}(a)
--\ref{fig:para-eusm}(h),
are taken from
Ref.~\cite{homma2025plb}.
Note that the boson-fermion interaction
strengths that are shown in
Figs.~1(e)--1(g) of \cite{homma2025plb}
with the notations $\Gamma$,
$\Lambda$, and $A$ in
the vertical axes
correspond to the parameters
denoted $\Gamma_0$,
$\Lambda_0$, and $A_0$
in the expressions \eqref{eq:coef_dyn}
--\eqref{eq:coef_mon}.
These differences in notations
do not at all affect the
final results and conclusions
given in the following,
and the calculations
in the present work and
in Ref.~\cite{homma2025plb}
are all consistent.

\begin{table}[htbp]
\centering
\caption{
Occupation probabilities
$v^2_j$ for the odd-$A$ nuclei
obtained from the HFB calculations
constrained to the deformation
$\beta = 0$.
The single-particle orbits
used in this study are
the proton $1h_{11/2}$ for the
Eu and La isotopes,
neutron $1i_{13/2}$ for the odd-$A$ Sm
and neutron $1h_{11/2}$ for the odd-$A$ Ba.
}
\begin{ruledtabular}
\begin{tabular}{cccccccc}
Nuclei & $v^2_{\pi h_{11/2}}$ & Nucleus & $v^2_{\nu i_{13/2}}$ & Nucleus & $v^2_{\pi h_{11/2}}$ & Nucleus & $v^2_{\nu h_{11/2}}$\\
\hline
$^{149}$Eu & 0.177 & $^{149}$Sm & 0.011 & $^{129}$La & 0.081 & $^{127}$Ba & 0.376 \\
$^{151}$Eu & 0.167 & $^{151}$Sm & 0.020 & $^{131}$La & 0.079 & $^{129}$Ba & 0.479 \\
$^{153}$Eu & 0.159 & $^{153}$Sm & 0.072 & $^{133}$La & 0.078 & $^{131}$Ba & 0.560 \\
$^{155}$Eu & 0.154 & $^{155}$Sm & 0.083 & $^{135}$La & 0.077 & $^{133}$Ba & 0.667 \\
$^{157}$Eu & 0.150 & $^{157}$Sm & 0.138 & & & & \\
$^{159}$Eu & 0.145 & $^{159}$Sm & 0.192 & & & & \\
\end{tabular}        
\end{ruledtabular}
\label{tab:vv}
\end{table}

We show from Figs.~\ref{fig:para-eusm}(a)
to \ref{fig:para-eusm}(d)
the IBM parameters 
for the even-even Sm isotopes
derived from the PEC-mapping procedure.
We can observe gradual evolution
of these parameters as functions
of $N$, confirming the empirical
evidence that the quadrupole
collectivity becomes stronger
with nucleon numbers in rare-earth
nuclei.
A notable change is found
in the systematic of the derived
parameter $\chi$, shown
in Fig.~\ref{fig:para-eusm}(c),
which rapidly decreases
from $N=92$ to 94 to a value
approximately close to the
SU(3) limit of the IBM,
$\chi=-\sqrt{7}/2\approx -1.32$ \cite{IBM}.
This irregularity at $N=94$
may have arisen from
numerical fitting, and thus
appears to be rather accidental.
Similar local behaviors
at $N=94$ are also obtained
in the parameters $\kappa$
and $C_{\rm B}$.
The present values of the IBM
parameters are compared with those
from earlier studies
using the generalized seniority scheme
\cite{scholten1985}
and with those from the
previous mapped IBM calculations
of Ref.~\cite{nomura2016qpt},
which employed the relativistic
EDF calculations to determine
the IBM Hamiltonian and $v^2_j$ values.
The present $\epsilon_d$ values
are consistent with the ones
in Refs.~\cite{scholten1985,nomura2016qpt}
in systematics, and the derived
$\chi$ parameters in this
study are similar to those
in Ref.~\cite{nomura2016qpt}.
However,
the $\kappa$ strengths adopted
here are significantly
different from the previous
two IBM studies.
This parameter seems to be
most sensitive to the microscopic
inputs from the fermionic calculations:
the $\kappa$ value of
\cite{nomura2016odd} was derived
with the inputs from the relativistic
EDF calculations, and the one
in \cite{scholten1985}
was obtained using the
schematic (surface delta)
shell model interaction.

In Figs.~\ref{fig:para-eusm}(e)--
\ref{fig:para-eusm}(h),
we show the IBFM parameters for $^{149-159}$Eu, which are
derived by using the formulas
\eqref{eq:IBFM_PES} and \eqref{eq:gen_pec}.
We can see that all these
parameters display a
drastic change from $N=88$ to 90.
The abrupt change of the parameters
reflects evolution of
nuclear structure in the odd-$A$ systems,
which takes place
along with that in the neighboring
even-even Sm cores.
What is worth a remark
is that the scale factor
$C_\mathrm{BF}$ for deformed odd-$A$
Eu nuclei with $N\geqslant90$ is chosen
to be unity, which means that 
the IBFM single-particle energies
are dictated by the same amount
of deformation as the
geometrical one in deformed nuclei.
Furthermore,
the present IBFM parameters, in most cases,
differ from those used
in the previous IBFM calculations
of Refs.~\cite{scholten1985,nomura2016odd}.
For instance, the present
$\Gamma_0$ and $\Lambda_0$ strengths
are larger than those in these earlier studies,
and the present $A_0$ strength
exhibits a stronger $N$ dependence.

The derived IBFM parameters for the
odd-$A$ Sm nuclei are depicted
in Figs.~\ref{fig:para-eusm}(i)--
\ref{fig:para-eusm}(l).
As in the case of the odd-$A$ Eu,
there appear discontinuous changes
of the parameters from $N=89$ to 91,
near which the even-even Sm undergo
a rapid shape phase transition.
We also compare our results
with the IBFM parameters
obtained in the previous
calculations of Ref.~\cite{nomura2016qpt}.
The present $\Gamma_0$ values
are of the same order of magnitude
as those in \cite{nomura2016qpt},
except for the nucleus $^{151}$Sm.
In the present study,
much smaller $\Lambda_0$ values
are obtained for the nuclei with
$N\leqslant89$ than those in
\cite{nomura2016qpt},
but those $\Lambda_0$ values
derived here for $^{153}$Sm
and $^{155}$Sm are consistent
with those in \cite{nomura2016qpt}.
The present monopole strengths
$A_0$ are generally larger in magnitude
than the ones in \cite{nomura2016qpt},
in which $A_0=0$ MeV for $N\geqslant89$.
Among the odd-$A$ systems
under investigation,
exchange strengths
chosen in the present work and
in Ref.~\cite{nomura2016qpt}
for the odd-$A$ Sm
turn out to be particularly large,
$\Lambda_0>10$ MeV.
This is due to the small
$v^2_{\nu i_{13/2}}$ values
(see Table~\ref{tab:vv}),
which according to the
formula \eqref{eq:coef_exc}
lead to vanishing $\beta_{i_{13/2},i_{13/2}}$
values.

Figures~\ref{fig:para-laba}(a)--
\ref{fig:para-laba}(d) show
the IBM parameters for the even-even
$^{128-134}$Ba nuclei, obtained from
the PEC-mapping procedure.
These IBM parameters exhibit a
more gradual evolution than 
those for the Sm isotopes.
This is consistent with the empirical
finding that the shape phase transition
occurs moderately in the mass
$A\approx130$ region.
In the $\gamma$-soft systems,
the potential energy surface
is approximately independent
of the $\gamma$ deformation,
which is characteristic of the
O(6) dynamical symmetry of the
IBM, and the $\chi$ parameter
in the quadrupole operator is
supposed to be close to zero.
One can indeed see
in Fig.~\ref{fig:para-laba}(c)
that the absolute value of $\chi$ for
each Ba isotope is generally
$\abs{\chi}<0.5$, which is much smaller
than the SU(3) limit,
$\abs{\chi}=\sqrt{7}/2\approx 1.32$.

One should also
notice in Figs.~\ref{fig:para-laba}(a)
and \ref{fig:para-laba}(b)
abrupt changes of the derived
energy $\epsilon_d$
and strength $\kappa$ from $N=76$ to 78.
These behaviors also reflect the
shape phase transition in the
Ba isotopes chain.
As one can see in Fig.~\ref{fig:pec-evenba}
the potential valley in
the PEC for $^{134}$Ba is much
shallower than that of the
neighboring $^{132}$Ba.
In terms of the IBM,
the shape transition is expected
to occur in the Ba isotopes from
the O(6) to U(5) limits near $N=78$,
and the nuclei $^{132}$Ba and $^{134}$Ba
have been considered to be
close to the O(6) limit.
The nucleus $^{134}$Ba was also
identified \cite{casten2000-E5}
as an empirical realization
of the E(5) critical-point symmetry
\cite{iachello2000-E5} of the
O(6)-U(5) QPT.

The IBM parameters obtained from
the mapping from the relativistic
SCMF calculations in
Ref.~\cite{nomura2017odd-1} are given
in Figs.~\ref{fig:para-laba}(a)
--\ref{fig:para-laba}(d).
They show similar systematic to those
of the present parameter values.
However, the values of the $\chi$
parameter for $^{128}$Ba and $^{130}$Ba
in the present work are larger
in magnitude than those
used in \cite{nomura2017odd-1}.
The difference indicates that the
Skyrme-HFB PEC is steeper in $\gamma$
deformation than the
relativistic one.
This conclusion may be, however,
altered if the triaxial degree
freedom is taken into account
in the Skyrme HFB calculation.

The IBFM Hamiltonian parameters
for the odd-$A$ $^{129-135}$La are
shown in Figs.~\ref{fig:para-laba}(e)
--\ref{fig:para-laba}(h).
Most of the derived
parameters do not show any strong
dependence on $N$.
These behaviors conform to
the gradual shape evolution
in the neighboring even-even Ba.
Figures \ref{fig:para-laba}(e)
--\ref{fig:para-laba}(g) give
the boson-fermion strengths
of Ref.~\cite{nomura2017odd-1},
which were fitted to experiment.
The $\Gamma_0$ and $\Lambda_0$
strengths of \cite{nomura2017odd-1}
are much smaller than those
derived in the present work.
The derived $\Lambda_0$, $A_0$,
and $C_{\rm BF}$ values for
the odd-$A$ $^{127-133}$Ba isotopes,
shown Figs.~\ref{fig:para-laba}(i)
--\ref{fig:para-laba}(l), exhibit
a gradual change with $N$,
but the adopted $\Gamma_0$ value for
$^{129}$Ba appears to be
anomalously large.
This is due to the HFB occupation
probability $v^2_{\nu h_{11/2}}=0.479$
for this nucleus (see Table~\ref{tab:vv}),
which is rather close to 0.5.
With $v^2_{\nu h_{11/2}}\approx0.5$,
meaning that the $1h_{11/2}$
neutron orbit is
approximately half filled,
the coefficient $\beta_{h_{11/2},h_{11/2}}$
in \eqref{eq:coef_exc}
becomes so negligibly small
that the large $\Gamma_0$
value is required.
The strength parameters
$\Gamma_0$, $\Lambda_0$, and
$A_0$ obtained in
Ref.~\cite{nomura2017odd-1}
exhibit behaviors that in general
differ from those in the
present work.
It is, however, also noted that
in Ref.~\cite{nomura2017odd-1}
for the odd-$N$ Ba nucleus with
mass $A$ the corresponding boson core
was taken to be the even-even Ba nucleus
with mass $A-1$, which means
that the boson core
for the odd-$N$ Ba nucleus
in Ref.~\cite{nomura2017odd-1}
was taken to be different
from that in
the present calculation.

\begin{figure*}[htbp]
\centering
\includegraphics[width=.9\linewidth]{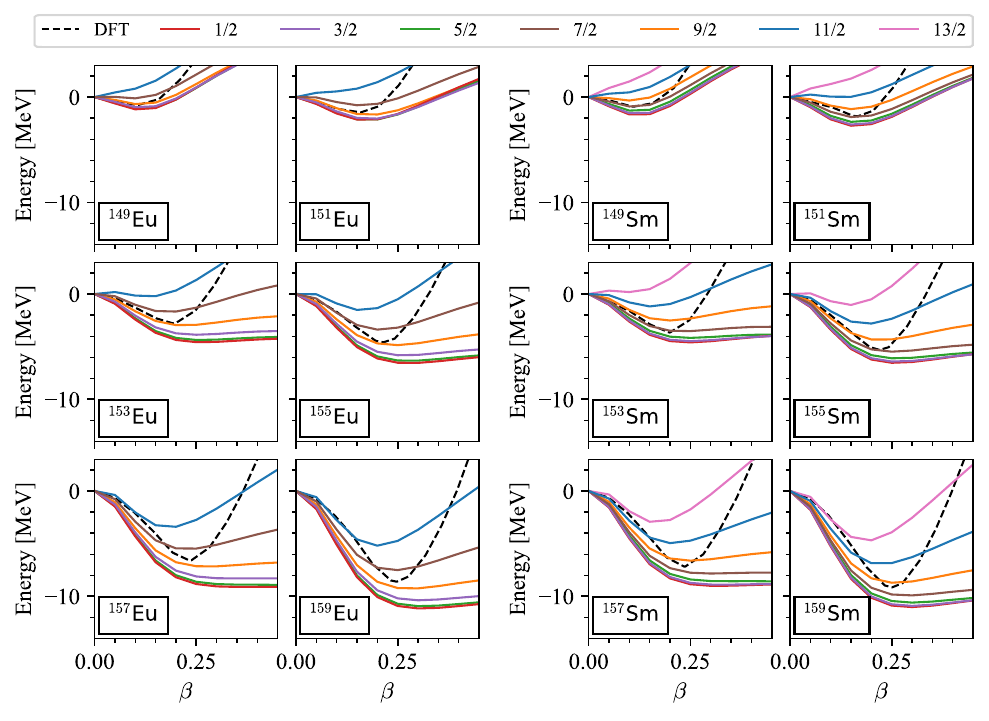}
\caption{Potential energy curves
for the odd-$A$ nuclei,
$^{149-159}$Eu and $^{149-159}$Sm,
calculated in the IBFM (solid curves)
for the quantum number
$K={1/2},3/2,\ldots,11/2$ (for Eu)
and $K=1/2,3/2,\ldots,{13/2}$ (for Sm).
The dashed curves represent
the Skyrme-HFB PECs.}
\label{fig:pec-eusm}
\end{figure*}

\begin{figure*}[htbp]
\centering
\includegraphics[width=.9\linewidth]{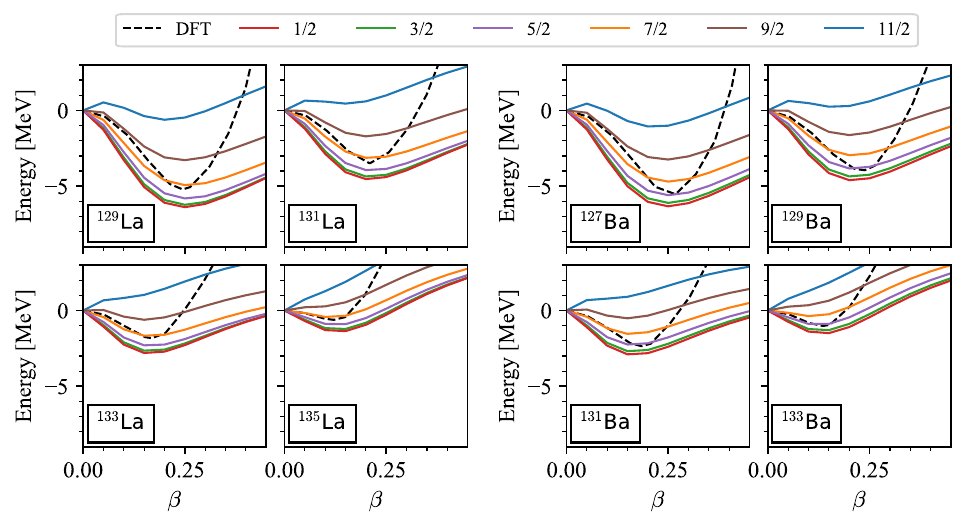}
\caption{Same as Fig.~\ref{fig:pec-eusm},
but for $^{129-135}$La and $^{127-133}$Ba.}
\label{fig:pec-laba}
\end{figure*}

\subsection{Potential energy curves
in odd-$A$ systems}\label{subsec:pes-o}

We show in Fig.~\ref{fig:pec-eusm}
the HFB and IBFM PECs \eqref{eq:gen_pec} for
the odd-$A$ nuclei $^{149-159}$Eu
and $^{149-159}$Sm,
corresponding to the different values
of the $K$ quantum number, i.e.,
$K = 1/2,\ 3/2,\ \dots,\ 11/2$
and
$K = 1/2,\ 3/2,\ \dots,\ 13/2$,
respectively.
Because the IBFM PECs should be
obtained for each $K$ value,
it is not possible to
compare them directly with the HFB PEC.
Nevertheless, most of the IBFM PECs
with different $K$ values show 
similar behavior to the HFB PEC:
the potential becomes deeper as $N$ increases,
and its minimum shifts from
$\beta_e = 0.10$ (0.11) to 0.25 (0.25) 
in Eu (Sm) isotopes.
We can infer from these behaviors of
both the IBFM and HFB PECs
a signature of the
phase transition in the odd-$A$ Eu
isotopes at the mean-field level.
Similar observations seem to hold
for the results for the
odd-$A$ Sm isotopes,
which are shown on the right-hand
side of Fig.~\ref{fig:pec-eusm}.

Figure~\ref{fig:pec-laba} shows
the IBFM PECs for both the odd-$A$
La and Ba isotopes,
plotted for each $K$ quantum number,
$K = 1/2,3/2,\ldots,11/2$,
and the corresponding Skyrme-HFB PECs.
The IBFM PECs reproduce the trend
of the HFB PECs, namely, as $N$ increases
the potentials become 
shallower, and location of the
equilibrium minimum shifts
closer to the origin $\beta=0$.
Given that these nuclei are
expected to be $\gamma$ soft in nature,
for a more accurate analysis
it is necessary to study the behaviors of
the energy surfaces in both the
axial $\beta$ and
triaxial $\gamma$ deformations.

\begin{figure*}[htbp]
\centering
\includegraphics[width=\linewidth]{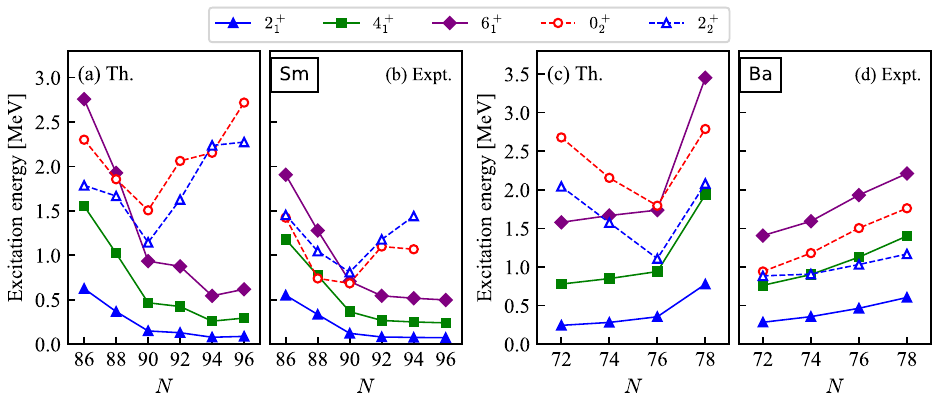}
\caption{
Calculated and experimental \cite{data}
low-energy spectra
for the even-even nuclei $^{148-158}$Sm
and $^{128-134}$Ba.}
\label{fig:ene-even}
\end{figure*}

\begin{figure*}[htbp]
\centering
\includegraphics[width=\linewidth]{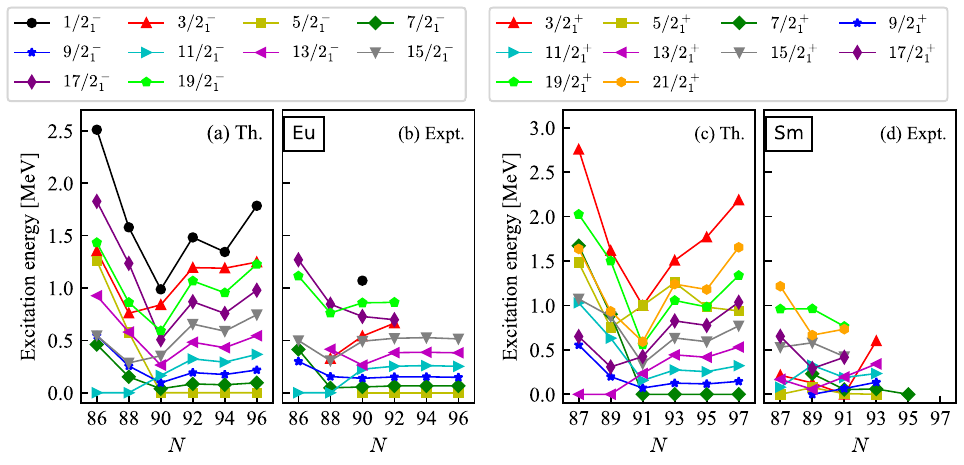}
\caption{Calculated and experimental \cite{data}
low-energy spectra of odd-$A$ Eu and Sm.}
\label{fig:ene-eusm}
\end{figure*}

\begin{figure*}[htbp]
\centering
\includegraphics[width=\linewidth]{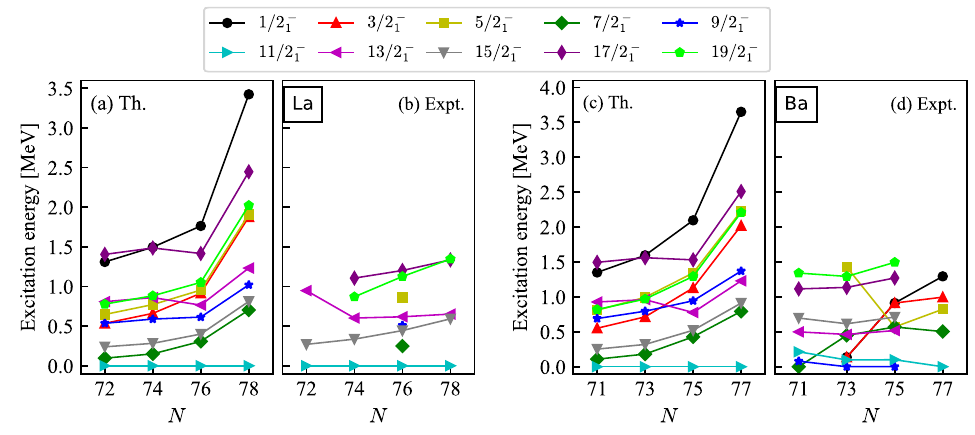}
\caption{Same as the caption to
Fig.~\ref{fig:ene-eusm},
but for odd-$A$ La and Ba.}
\label{fig:ene-laba}
\end{figure*}

\subsection{Energy spectra} \label{subsec:energy}

Low-energy spectra for the
even-even core nuclei, $^{148-158}$Sm
and $^{128-134}$Ba, are shown
in Fig.~\ref{fig:ene-even}.
The IBM calculation overall
reproduces the observed energy
spectra well.
The ratio $R_{4/2}$ of the $4^+_1$
to $2^+_1$ excitation energies
serves as an indicator of the
shape transition.
The measured $R_{4/2}$ ratios
are 2.15, 2.32, 3.00, 3.30, 3.29,
and 3.29 for $^{148-158}$Sm,
respectively.
The $R_{4/2}$ value for
$^{148}$Sm is close to the vibrational
U(5) limit ($R_{4/2}=2$),
and those for $^{154,156,158}$Sm
are close to the value in
the rotational SU(3)
($R_{4/2}=10/3$) limit.
The present calculation reproduces
this systematic,
with the calculated
$R_{4/2}$ values being 2.49, 2.79, 3.13,
3.29, 3.32, and 3.34
for $^{148-158}$Sm, respectively.
The overestimates of $R_{4/2}$ in
the nearly spherical, U(5), regime
are due to the fact that
the HFB PECs exhibit
a rather large deformation,
$\beta_e = 0.10$ and 0.15
for $^{148,150}$Sm, respectively,
and the potential is very steep.
These features of the PECs
are supposed to be reflected
in the final IBM spectra,
which have the $R_{4/2}$ ratios
significantly larger than 2.0.

One can see
from Figs.~\ref{fig:ene-even}(c)
and \ref{fig:ene-even}(d) that
the mapped IBM calculations
reproduce the overall 
systematic of the yrast levels in the
even-even Ba nuclei with
$72\leqslant N \leqslant76$,
but the predicted $0^+_2$
and $2^+_2$ energies exhibit
decreasing patterns with $N$
in contrast to the observed ones,
which gradually increase in energy.
For $^{128}$Ba and $^{130}$Ba,
the $0^+_2$ and $2^+_2$
excitation energies are also
overestimated in the
present calculation,
and the level structure of
these nuclei rather resembles
that of rotational spectra.
The mapped IBM qualitatively
reproduces the increases of the
$2^+_1$, $4^+_1$, $0^+_2$ and $2^+_2$
energy levels from
$N=76$ to 78.
The order of levels
is reasonably reproduced
for $^{132}$Ba and $^{134}$Ba,
in which the $2^+_2$ levels
are lying close to the
$2^+_1$ ones, a signature
of $\gamma$ softness.

Figure~\ref{fig:ene-eusm}(a)
shows the predicted low-energy spectra
of negative-parity yrast states 
in odd-$A$ $^{149-159}$Eu,
obtained from the IBFM,
with the Hamiltonian parameters
specified by the EDF inputs
using the procedure described
in Sec.~\ref{sec:model}.
One of the empirical signatures
of the possible QPTs in odd-$A$ nuclei
is a change in the spin of the 
lowest-energy state of a given parity
at a particular nucleon number
\cite{petrellis2011}.
In the present case,
the observed energy spectra, shown in
Fig.~\ref{fig:ene-eusm}(b),
exhibit such a feature, as
the spin $I$ of the lowest-lying
negative-parity
state changes from $11/2^-$
to $5/2^-$ at $N=90$. 
The mapped IBFM consistently
reproduces this trend.
In addition, the $11/2^-$ yrast
states in $^{149}$Eu and $^{151}$Eu
are mostly accounted
for by the proton $1h_{11/2}$ single-particle
configuration coupled to the
weakly-deformed even-even core,
and the bands based on the
${11/2}^-_1$ state show the
$\Delta I=2$ sequence of levels,
connected by dominant $E2$ transitions.
The level structure clearly changes
at $N=90$, since the yrast levels
in those Eu nuclei with $N \geqslant 90$
rather resemble a rotational band
following a $\Delta I = 1$ sequence.
The present mapped IBFM reasonably
reproduces these level structures,
but overestimates the lower-spin states,
e.g., with $I={1/2}^-$ and ${3/2}^-$.
These low-spin states could probably
arise from the correlations
that are not incorporated
in the present implementation
of the IBFM, such as
configuration mixing.

We also show in Fig.~\ref{fig:ene-eusm}
the predicted energy levels of the
positive-parity yrast states
in $^{149-159}$Sm.
These states arise from the coupling
of the neutron $1i_{13/2}$
single-particle state to the
even-even Sm cores in the IBFM.
Similarly to the odd-$A$ Eu case,
the mapped IBFM predicts
the spin $I$ of the lowest-lying
positive-parity states
in $^{149}$Sm and $^{151}$Sm
to be ${13/2}^+$,
and the subsequent $\Delta I=2$ bands.
For those odd-$A$ Sm nuclei
with $N\geqslant91$,
the calculation suggests the
$\Delta I=1$ rotational-like
bands built on the ${7/2}^+$ states.
The spectroscopic data for the
odd-$A$ Sm, however, show a much
more complicated level structure than
those for the odd-$A$ Eu.
In particular, the lowest-lying
level structure in transitional
nuclei such as those with $N=89$ and 91
is characterized by coexistence of
several decoupled, $\Delta I=2$,
and strongly-coupled, $\Delta I=1$,
bands lying quite close in energy.
The present IBFM calculation
is not able to reproduce full details
of the observed energy spectra
in the odd-$A$ Sm,
including the behavior of the
${5/2}^+_1$ level, which
appears in the vicinity of
the lowest-lying state experimentally.

In Fig.~\ref{fig:ene-laba}
the predicted excitation 
energies of negative-parity yrast
states in $^{129-135}$La and $^{127-133}$Ba
are compared with the experimental data.
Our calculation suggests the
lowest-lying negative-parity
state to be ${11/2}^-$ for all
the odd-$A$ La, and produces
a gradual increase
of the $I={15/2}^-$ level with $N$
consistently with experiment.
The predicted energy spectra
for the odd-$A$ La isotopes show
an overall gradual increase
as $N$ approaches the
neutron major shell closure $N=82$.
Note that the overestimate of
the energy levels of $^{135}$La
could be attributed
to the irregular behaviors
of some of the IBM parameters
at $N=78$, in particular,
the large $d$-boson energy,
$\epsilon_d$, obtained for $^{134}$Ba
[see Figs.~\ref{fig:para-laba}(a)
--\ref{fig:para-laba}(d)].

Patterns of the low-energy
negative-parity spectra
for the odd-$A$ Ba, shown on
the right-hand side of
Fig.~\ref{fig:ene-laba}, look
similar to those in the odd-$A$ La.
Namely, in both isotopic chains
the $I={11/2}^-$ state is the
lowest-lying state, and the whole
energy spectrum becomes more
stretched as $N$ increases.
The predicted energy levels, except
for that of the ${9/2}^-$ state, are
in most cases consistent with the
experimental counterparts.
The evolution of the energy levels
in the odd-$A$ La and Ba appears to
occur only gradually,
and is not as pronounced
as in the case of the rare-earth
region.

\begin{figure*}[htbp]
\centering
\includegraphics[width=\linewidth]{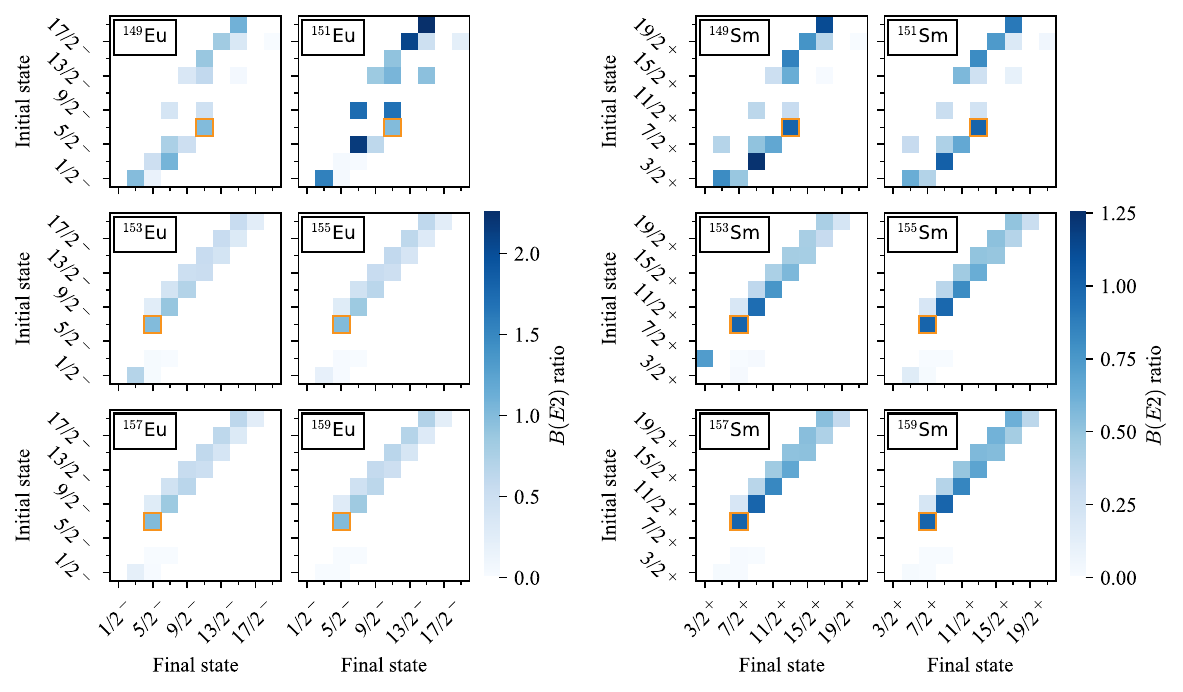}
\caption{
Predicted
$B(E2)$ values in $^{149-159}$Eu and
$^{149-159}$Sm, normalized
with respect to
those $E2$ transitions from
the second-lowest-energy state to the
lowest-energy state of a given
parity (marked by the thick square
in each plot).
}
\label{fig:e2-eusm}
\end{figure*}

\begin{figure*}[htbp]
\centering
\includegraphics[width=\linewidth]{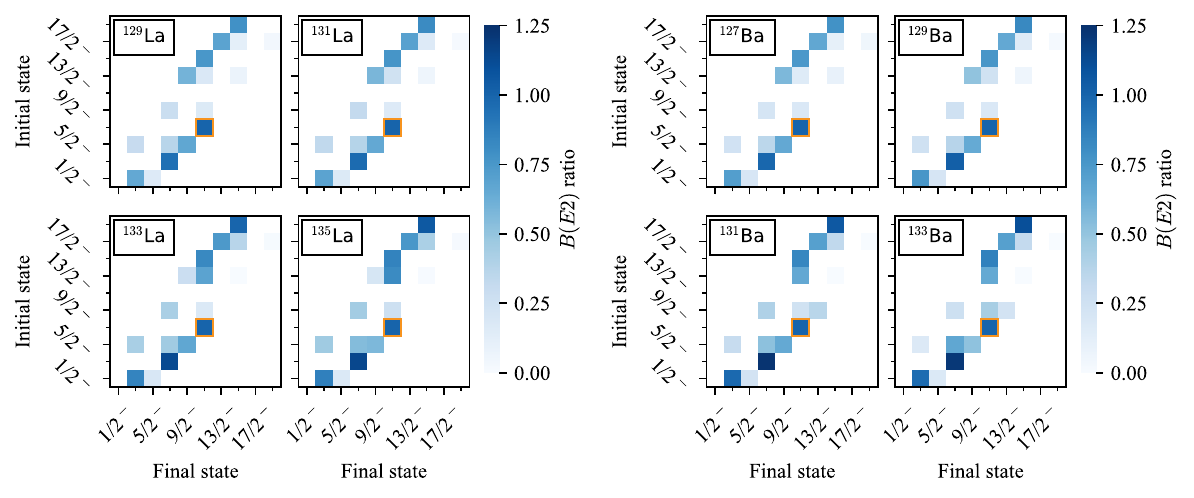}
\caption{
Same as the caption to
Fig.~\ref{fig:e2-eusm}, but for
$^{129-135}$La and $^{127-133}$Ba.}
\label{fig:e2-laba}
\end{figure*}

\subsection{E2 transitions} \label{subsec:e2}

$E2$ transition properties are calculated
by using the transition operator
that is given as
\begin{equation}
\hat{T}^\mathrm{(E2)}_\mathrm{BF} = \hat{T}^\mathrm{(E2)}_\mathrm{B} + \hat{T}^\mathrm{(E2)}_\mathrm{F} \; ,
\label{eq:e2op}
\end{equation}
where the first and second terms
represent the contributions from
the even-even boson core and odd fermion,
and are given, respectively, as
\begin{align}
&\hat{T}^\mathrm{(E2)}_\mathrm{B}
=e_\mathrm{B}\hat{Q}
\label{eq:e2-b}\\
&\hat{T}^\mathrm{(E2)}_\mathrm{F}
=-e_\mathrm{F}\frac{1}{\sqrt{5}}\gamma_{jj}
[a_j^\+\cross\tilde{a}_j]^{(2)} \; .
\label{eq:e2-f}
\end{align}
$e_\mathrm{B}$ and $e_\mathrm{F}$
are bosonic and fermionic
effective charges, respectively.
The operator $\hat{Q}$ in \eqref{eq:e2-b}
is the same quadrupole operator
with the same $\chi$ parameter
as those in the IBM Hamiltonian
\eqref{eq:hb}.
The factor $\gamma_{jj}$ is
defined in \eqref{eq:coef_dyn}.
In most IBM calculations,
the effective charges
$e_\mathrm{B}$ are usually determined
so as to reproduce the experimental
$B(E2)$ values, and are also
taken to be constant in a given
isotopic chain.
Here we assume $e_{\rm B}$ to
be dependent on nucleon numbers,
and determine its values
by means of the prescription
of Ref.~\cite{rudigier} relating
the $E2$ reduced matrix element
to the deformation $\beta$ as
\begin{equation}
\beta_e = \frac{4\pi}{3ZR_0^2}\sqrt{\be 0 1 2 1} \; ,
\label{eq:beta-be2}
\end{equation}
where the $\beta_e$ value
is obtained from the HFB PEC,
$\be 0 1 2 1$ is calculated
in the IBM, and
$R_0=\SI[parse-numbers=false]{1.2A^{1/3}}{fm}$. 
The above formula gives the value
of $e_{\rm B}$:
\begin{equation}
e_\mathrm{B}
=\frac{3ZR_0^2\beta_e}{4\pi|\mel*{2_1^+}{|\hat{Q}|}{0_1^+}|} \; .
\label{eq:eb}
\end{equation}
The resultant $e_{\rm B}$'s for
each even-even and odd-$A$
nucleus are given in Tables~\ref{tab:eb-eusm}
and \ref{tab:eb-laba}.
The proton and neutron effective
charges,
$e_\mathrm{F}=1.5\,e\mathrm{b}$
and
$e_\mathrm{F}=0.5\,e\mathrm{b}$,
respectively,
which are widely used in the nuclear
shell model calculations,
are adopted for the fermionic
operator $\hat T^{(E2)}_{\rm F}$
\eqref{eq:e2-f}.

\begin{table}[htbp]
\centering
\caption{Bosonic effective charges
in $e$b units
for the even-even $^{148-158}$Sm
isotopes adopted in
the present calculation,
which are used for the odd-$A$ Eu
and Sm.}
\begin{ruledtabular}
\begin{tabular}{ccccccc}
$N$ & 86 & 88 & 90 & 92 & 94 & 96 \\
\hline
$e_\mathrm{B}$ & 0.068 & 0.085 & 0.092 & 0.092 & 0.124 & 0.145
\end{tabular}        
\end{ruledtabular}
\label{tab:eb-eusm}
\end{table}

\begin{table}[htbp]
\centering
\caption{Same as the caption
to Table.~\ref{tab:eb-eusm},
but for the even-even $^{128-134}$Ba,
and the neighboring odd-$A$ La and Ba isotopes.}
\begin{ruledtabular}
\begin{tabular}{ccccc}
$N$ & 72 & 74 & 76 & 78 \\
\hline
$e_\mathrm{B}$ & 0.115 & 0.113 & 0.102 & 0.080 
\end{tabular}        
\end{ruledtabular}
\label{tab:eb-laba}
\end{table}

The calculated $\be 2 1 0 1$ values
for the even-even nuclei $^{148-154}$Sm,
with $e_{\rm B}$ being determined
by using the formula \eqref{eq:eb},
are 17 Weisskopf units (W.u.),
34 W.u., 51 W.u.,
and 99 W.u.,
respectively.
These values are lower than those
of the experimental data \cite{data}
by a factor of 2 to 3:
$30.7\pm1.5$ W.u.,
$57.0\pm1.3$ W.u.,
$143.683\pm0.087$ W.u.,
and $177.3\pm1.8$ W.u.,
respectively.
This is due mainly to the too small
effective charges $e_\mathrm{B}$
derived for the even-even Sm nuclei
(cf. Table~\ref{tab:eb-eusm}).
The calculated $\be 2 1 0 1$
values for $^{128-134}$Ba are 
72 W.u., 51 W.u., 30 W.u., and 13 W.u.,
respectively, which compare rather well
the experimental values \cite{data},
$72\pm7$ W.u.,
$57.9\pm1.7$ W.u.,
$43\pm4$ W.u., and
$33.6\pm0.6$ W.u.,
respectively.

Figure~\ref{fig:e2-eusm} depicts
in a color map
the predicted $B(E2)$ values for the
transitions between the negative-parity
yrast states and between
the positive-parity yrast states
in the considered odd-$A$ Eu and Sm nuclei,
respectively.
These $B(E2)$ values are
normalized with respect to those
of the $E2$ transitions
from the second-lowest- to lowest-energy
states of a given parity:
$\ben 7 1 {11} 1$ = 23 W.u. and 22 W.u.
for $^{149,151}$Eu,
$\ben 7 1 5 1=$ 68 W.u., 101 W.u., 307 W.u.,
and 417 W.u. for $^{153-159}$Eu, 
$\bep 9 1 {13} 1 =$ 21 W.u., and 59 W.u. 
for $^{149,151}$Sm, and 
$\bep 9 1 7 1 =$ 56 W.u., 86 W.u., 272 W.u., 
and 376 W.u. for $^{153-159}$Sm, respectively.

In Fig.~\ref{fig:e2-eusm},
for $^{149}$Eu and $^{151}$Eu
the $B(E2)$'s for the $\Delta I=2$
transitions, $\bejn {19} {1} {15} {1}$
and $\bejn {15} {1} {11} {1}$, are large
compared with those of the $\Delta I=1$
$E2$ transitions.
There appear to be
certain degrees of mixing particularly
among those states with mid-spin,
i.e., $I\approx{9/2}$,
which exhibit dominant
$\Delta I=1$ transitions.
For $^{153-159}$Eu, the dominant $B(E2)$ values
between states whose spins differ by 1
show the typical rotational band following
the $\Delta I = 1$ systematic.
Similar $B(E2)$ systematic is obtained
for the odd-$A$ Sm isotopes,
which is presented on the
right-hand side of Fig.~\ref{fig:e2-eusm}.

We compare in Table~\ref{tab:e2}
predicted $B(E2)$ and
spectroscopic quadrupole moments
$Q(I)$ for the odd-$A$ Eu and Sm
with the experimental data \cite{data}.
There are only a few experimental
data to compare for the
negative-parity states in odd-$A$ Eu.
For $^{151}$Eu, only the lower limit
of $\ben 9 1 7 1\geqslant 70$ W.u.
is known experimentally.
Our calculation gives $\ben 9 1 7 1=37.9$
W.u. for this nucleus,
underestimating the experimental
lower limit by a factor of 2.
This discrepancy is considered
to be due to the too small
bosonic effective
charge $e_\mathrm{B}$ for the
boson core nucleus $^{150}$Sm,
which is here derived by the
relation in \eqref{eq:eb} using
the $\beta_e$ value provided by
the Skyrme-HFB calculation.
The same seems to be
true for the
$\ben {17} 1 {13} 1$ for $^{153}$Eu
and $\bep 5 1 9 1$ for $^{151}$Sm.
The IBFM gives a smaller
quadrupole moment $Q({3/2}^+_1)$
than the observed one,
but the correct sign is obtained.

Figure~\ref{fig:e2-laba} gives
a plot similar to Fig.~\ref{fig:e2-eusm},
but in the case of the odd-$A$ La
and Ba nuclei.
The $B(E2)$ values shown in
Fig.~\ref{fig:e2-laba} are
those normalized with respect to
$\ben 7 1 {11} 1 =$ 140 W.u., 93 W.u., 
39 W.u., and 17 W.u. for $^{129-135}$La, 
and $\ben 7 1 {11} 1 =$ 138 W.u., 87 W.u.,
34 W.u., and 15 W.u.
for $^{127-133}$Ba, respectively.
We can see that in general all the odd-$A$
La and Ba isotopes
studied here indicate dominant
$\Delta I = 2$ $E2$ transitions.
As in the case of the weakly deformed
odd-$A$ Eu and Sm with mass $A\leqslant151$,
shown in Fig.~\ref{fig:e2-eusm},
there appear to be certain mixing
among low- and mid-spin states.
The predicted $B(E2)$ values
for some transitions in the odd-$A$ La
by the mapped IBFM are compared with
the measured
values in Table~\ref{tab:e2}.
The IBFM reproduces the observed
$B(E2)$ values fairly well.
The $B(E2)$ values obtained from
the IBFM for $^{129}$Ba
are more or less
consistent with the data.
The present model calculation,
however, gives the negative
spectroscopic
quadrupole moments for $^{127}$Ba,
$^{131}$Ba, and $^{133}$Ba
at variance with the experimental data.
This arises because the
underlying Skyrme-HFB PECs
for the even-even Ba cores
show a rather pronounced prolate deformation
(see Fig.~\ref{fig:pec-laba}),
and the resulting parameter
$\chi$ is negative in sign.

The deficiencies in descriptions
of the electromagnetic transition
properties in the odd-$A$ nuclei,
including the $B(E2)$
values for Sm and Eu being underestimated,
and the incorrect sign for the
spectroscopic quadrupole moments
in La and Ba,
could be naturally attributed to
the microscopic inputs provided
by the mean-field calculations,
since in the present theoretical
framework they influence significantly
the boson effective charges \eqref{eq:eb}
and the values of the IBFM
parameters, in particular,
the parameter $\chi$
in the boson-core Hamiltonian,
which determines, to a large
extent, the sign of the
quadrupole moments.
In addition, some interaction
terms in the IBM Hamiltonian
that potentially have impacts
on the electromagnetic transitions
may have been missing in
the present calculation.
For instance, as pointed out,
the inclusion of the triaxial degree of
freedom and cubic interactions
in the mapping procedure
could influence the predicted values
of the $E2$ transition properties
in $\gamma$-soft nuclei.

\begin{table}[htbp]
\centering
\caption{
Predicted
$B(E2)$ values (in W.u.)
and $Q(I)$ (in $e$b) moments of the odd-$A$
Eu, Sm, La, and Ba isotopes in comparison with the
experimental data taken from
Refs.~\cite{data,stone2005}.}
\begin{ruledtabular}
\begin{tabular}{cccc}
Nucleus & Properties & IBFM & Expt. \\
\hline
$^{151}$Eu & $\ben 9 1 7 1$ & 37.9 & $>70$ \\
$^{153}$Eu & $\ben {17} 1 {13} 1$ & 36.6 & $196^{+36}_{-31}$ \\
$^{151}$Sm & $\bep 5 1 9 1$ & 60.1 & $170\pm30$ \\
$^{153}$Sm & $Q(3/2^+_1)$ & $+0.497$ & $+1.30\pm0.12$ \\
$^{129}$La & $\ben {15} 1 {11} 1$ & 105 & $107\pm5$ \\
           & $\ben {19} 1 {15} 1$ & 109 & $100\pm15$ \\
$^{131}$La & $\ben {15} 1 {11} 1$ & 70.5 & $87\pm3$ \\
           & $\ben {19} 1 {15} 1$ & 75.0 & $87\pm10$ \\
$^{127}$Ba & $Q(7/2^-_1)$ & $-1.03$ & $+1.62\pm0.13$ \\
$^{129}$Ba & $\ben {15} 1 {11} 1$ & 66.4 & $54.5\pm2.3$ \\
           & $\ben {19} 1 {15} 1$ & 71.9 & $90\pm40$ \\
$^{131}$Ba & $Q(9/2^-_1)$ & $-0.534$ & $+1.46\pm0.13$ \\
$^{133}$Ba & $Q(11/2^-_1)$ & $-0.632$ & $+0.89\pm0.07$
\end{tabular}        
\end{ruledtabular}
\label{tab:e2}
\end{table}

\section{summary and conclusions}  \label{sec:summary}

We have developed a novel EDF-based
collective model
for the quantitative and systematic
calculations of the spectroscopic properties
in odd-$A$ nuclei.
The parameters for the IBFM, which
provides spectroscopy of odd-$A$ nuclei,
have been completely determined by
using the solutions of the SCMF calculations
based on the nuclear EDF:
the boson-core Hamiltonian is
specified by mapping the
EDF-SCMF potential energy
onto the expectation value of the
IBM Hamiltonian;
the boson-fermion coupling constants
are derived by associating the
deformed single-particle
spectra in the intrinsic frame
of the IBFM with the SCMF
counterparts.
By extending the preceding
work of Ref.~\cite{homma2025plb},
we here studied systematically the
low-lying structure in the odd-$Z$
$^{149-159}$Eu and
odd-$N$ $^{151-159}$Sm nuclei,
in which the shape phase transitions
from nearly spherical to strongly deformed
are expected to occur,
and in the odd-$Z$
$^{129-135}$La and odd-$N$ $^{127-133}$Ba
nuclei in the region in which
$\gamma$-soft shapes are likely to emerge.

It was demonstrated that
the energy spectra of
the odd-$A$ nuclei as well as those
of the even-even core nuclei
were calculated only by references
to the microscopic inputs provided
by the nuclear EDF,
hence no phenomenological fit of the
model parameters to experiment
is required.
Our microscopic IBFM
reproduced quantitatively
the observed energy spectra
in the odd-$Z$ Eu and, in particular,
reproduced the change of
spin of the lowest-lying
negative-parity state near
$N=90$ as a signature of
the possible QPT.
Our calculation suggested
a gradual evolution of the energy
spectra with $N$ in the
odd-$A$ La and Ba that is consistent with
the experimental data, and indicates
a shape phase transition
from $\gamma$ soft to nearly
spherical regimes.
We further devised a way of
calculating the $E2$ transition
probabilities with the boson
effective charges determined without
directly comparing to the experimental
$B(E2)$ data.
This provides insights into the band
structure and electromagnetic
transition properties in odd-$A$
systems, for which the $B(E2)$
data are scarce.

These results demonstrate the validity
of the proposed method
in the general quadrupole collective
states,
namely, spherical vibrational U(5),
deformed rotational SU(3),
and $\gamma$-unstable O(6) limits,
in the presence of the odd fermion
in a single-$j$ orbit.
It was, however, also found
that the detailed energy-level structure
in odd-$N$ Sm and Ba isotopes
was not accurately described.
Hwever, we consider these results
for the odd-$N$ systems to be
rather reasonable,
given that the IBFM
calculations are here
not specifically adjusted to data,
but are based only on
the EDF inputs.
In addition,
a more complete description
of the $\gamma$-soft nuclei
should explicitly take into account
the triaxial degree of freedom
and subsequently the
cubic interactions in
the boson core Hamiltonian.
The EDF-IBFM framework in its present
stage is able to capture gross systematic
trends of the low-energy spectra
in the odd-$A$ $\gamma$-soft nuclei
that are consistent with experiment.

Since the formalism presented
in this article was limited to
single-$j$ cases,
a straightforward extension
would be to include multiple
orbits of normal parity in
the single-particle space.
For the positive-parity states
in odd-$A$ Eu, for instance,
there are plenty of spectroscopic
data including those for the
$E2$ properties,
which will be useful to test
the proposed methodology.
Another directions of the future study
concern inclusion of the neutron
and proton boson degrees of freedom,
i.e., an extension to the IBFM-2,
for a more realistic calculation.
This extension would be especially
crucial for predicting nuclear
processes such as $\beta$ decay.
In addition to the quadrupole mode,
the octupole degree of freedom
is expected to be relevant to
determine nuclear structure
in radioactive odd-$A$ nuclei in
certain mass regions of the nuclear
chart.
It is, therefore,
an interesting subject to incorporate
the octupole correlations
in our model and study their influences
on low-lying states in odd-$A$ nuclei.
The work along these directions
will be reported elsewhere.

\acknowledgements
The author MH acknowledges support
from JST SPRING, Grant No. JPMJSP2119.
The work of the author KN
has been supported by JSPS
KAKENHI Grant No. JP25K07293.

\bibliography{refs}

\end{document}